\documentclass[onecolumn,floatfix,superscriptaddress,showpacs,showkeys,nofootinbib,preprint]{revtex4}
\usepackage{epsfig}
\usepackage{amssymb,latexsym,amsmath}
\newcommand{\eq}[1]{\begin{align} #1 \end{align}}

\begin{document}
\title{
Multiplicity Fluctuations in Hadron-Resonance Gas}

\author{V.V. Begun}

\affiliation{
 Museo Storico della Fisica e Centro Studi e Ricerche
Enrico Fermi}
\affiliation{
 Bogolyubov Institute for Theoretical Physics, Kiev, Ukraine}

\author{M.I. Gorenstein}
\affiliation{
 Bogolyubov Institute for Theoretical Physics, Kiev, Ukraine}
\affiliation{Frankfurt Institute for Advanced Studies, Frankfurt,
Germany}
\author{M. Hauer}
\affiliation{University of Cape Town, South Africa}
\author{V.P. Konchakovski}
\affiliation{
 Bogolyubov Institute for Theoretical Physics, Kiev, Ukraine}
\author{O.S. Zozulya}
\affiliation{
 Bogolyubov Institute for Theoretical Physics, Kiev, Ukraine}
\affiliation{
 Utrecht University, Utrecht, The Netherlands}

\begin{abstract}
The  charged hadron multiplicity fluctuations are considered in
the canonical ensemble. The microscopic correlator method is
extended to include three conserved charges: baryon number,
electric charge and strangeness. The analytical formulae are
presented that allow to include resonance decay contributions to
correlations and fluctuations. We make the predictions for  the
scaled variances of negative, positive and all charged hadrons in
the most central Pb+Pb (Au+Au) collisions for different collision
energies from SIS and AGS to SPS and RHIC.

\end{abstract}



\maketitle

\section{Introduction}

The statistical models have been successfully used to describe the
data on hadron multiplicities in relativistic nucleus-nucleus
(A+A) collisions (see, e.g., Ref.\,\cite{stat-model,FOC,FOP} and
recent review \cite{BMST}). The applications of the statistical
model to elementary reactions and/or to rare particles production
have stimulated an investigation of the
relations between different statistical ensembles.
In A+A collisions one prefers to use the grand canonical ensemble
(GCE) because it is the most convenient one from the technical
point of view. The canonical ensemble (CE)
\cite{ce-a,ce,ce-b,ce-c,ce-d,ce-e} or even the microcanonical
ensemble (MCE) \cite{mce} have been used in order to describe the
$pp$, $p\bar{p}$ and $e^+e^-$ collisions when a small number of
secondary particles is produced.  At these conditions the
statistical systems are far away from the thermodynamic limit, the
statistical ensembles are not equivalent, and the exact charge, or
both energy and charge conservation, laws have to be taken into
account. The CE suppression effects for particle multiplicities
are well known in the statistical approach to hadron production,
for example, the suppression in a production of strange hadrons
\cite{ce-c}, antibaryons \cite{ce-d}, and charmed hadrons
\cite{ce-e}  when the total numbers of these particles are small
(smaller than or equal to 1). The different statistical ensembles
are not equivalent for small systems. When the system volume
increases, $V\rightarrow\infty$, the average quantities in the
GCE, CE and MCE become equal to each other, i.e., all statistical
ensembles are thermodynamically equivalent.

The fluctuations in high energy nuclear collisions (see, e.g.,
Refs.~\cite{fluc1,fluc2,fluc3,fluc4,fluc5,step1,Jeon-Koch,fluc6,fluc7,fluc7a,fluc7b,fluc8,MGMG,KGB}
and references therein) reveal  new physical information and can
be closely related to the phase transitions in the QCD matter. The
particle number fluctuations for relativistic systems in the CE
were calculated for the first time in Ref.~\cite{ce-fluc} for the
Boltzmann ideal gas with net charge equal to zero. These results
were then extended to quantum statistics and non-zero net charge
in the CE \cite{ce2-fluc,ce3-fluc,bec,bose-fluc} and to the MCE
\cite{mce-fluc,mce2-fluc}, and compared to the corresponding
results in the GCE (see also Refs.~\cite{turko,Hauer}). Expressed
in terms of the scaled variances, the particle number fluctuations
have been found to be suppressed in the CE and MCE comparing to
the GCE. This suppression survives in the limit
$V\rightarrow\infty$, so the thermodynamical equivalence of all
statistical ensembles refers to the average quantities, but is not
applied to the scaled variances of particle number fluctuations.

The aim of the present paper is to extend a microscopic correlator
method to treat the hadron-resonance gas within the CE
formulation. In Section \ref{sec-QBS} we calculate the microscopic
correlators in relativistic quantum gas. This allows one to take
into account Bose and Fermi effects, as well as an arbitrary
number of the conserved charges in the CE. We also argue that the
microscopic correlator approach gives the same results as the
explicit saddle point CE calculations \cite{ce3-fluc,bec} in the
large volume limit $V\rightarrow\infty$. In Section \ref{sec-res}
we define the generating function to include the effects of
resonance decays. This gives the {\it analytical} expressions for
resonance decay contributions to the particle correlations and
fluctuations within the CE and MCE. In Section \ref{sec-HG} we
calculate and make the predictions for the scaled variances of
negatively, positively and all charged particles in central Pb+Pb
(Au+Au) collisions along the chemical freeze-out line at different
collision energies. Section \ref{sec-summary} presents our summary
and conclusions. Some details of the calculations are given in
Appendix.

\section{Exact charge conservations in statistical systems}\label{sec-QBS}

\subsection{CE Microscopic Correlator }

Let us consider the fluctuations in the ideal relativistic gas
within the CE. Our primary interest is to include different types
of hadrons, while keeping exactly fixed the global electric (Q),
baryon (B), and strange (S) charges of the statistical system.
The system of non-interacting Bose or Fermi particles of species
$i$ can be characterized by the occupation numbers $n_{p,i}$ of
single quantum states labelled by momenta $p$. The occupation
numbers run over $n_{p,i} = 0,\, 1$ for fermions and $n_{p,i} =
0,\, 1, 2, \ldots$ for bosons. The GCE average values and
fluctuations of $n_{p,i}$ equal the following \cite{lan}:
 \eq{
 \langle n_{p,i} \rangle
 ~& = ~\frac {1} {\exp \left[\left(\sqrt{p^{2}+m_i^{2}}~-~ \mu_i\right) / T\right]
 ~-~ \gamma_i}~, \label{np-aver}
 \\
 v^{ 2}_{p,i}
 ~& \equiv~ \langle\Delta n_{p,i}^2\rangle
 ~ \equiv~ \langle \left( n_{p,i}-\langle
 n_{p,i}\rangle\right)^2\rangle
 ~=~ \langle n_{p,i}\rangle
\left(1 + \gamma_i \langle n_{p,i} \rangle\right)~.\label{np-fluc}
  }
In Eq.~(\ref{np-aver}), $T$ is the system temperature, $m_i$ is
the mass of $i$-th particle species, $\gamma_i$ corresponds to
different statistics ($+1$ and $-1$ for Bose and Fermi,
respectively, and $\gamma_i=0$ gives the Boltzmann approximation),
and chemical potential $\mu_i$ equals:
\eq{ \mu_i~=~q_i~\mu_Q~+~b_i~\mu_B~+~s_i~\mu_S ~,\label{chempot}}
where $q_i,~b_i,~s_i$ are the electric charge, baryon number and
strangeness of particle of species $i$, respectively, while
$\mu_Q,~\mu_B,~\mu_S$ are the corresponding chemical potentials
which regulate the average values of these global conserved
charges in the GCE.

The average number of particles of species $i$, the number of
positive, negative, and all charged particles are equal:
 \eq{\label{Ni-gce}
 \langle N_i\rangle \;& \equiv\; \sum_p \langle n_{p,i}\rangle
 \;=\; \frac{g_i V}{2\pi^{2}}\int_{0}^{\infty}p^{2}dp\; \langle
 n_{p,i}\rangle\;,\\
  \langle N_{+}\rangle \;&=\; \sum_{i,q_i>0} \langle N_i\rangle\;,
 ~~~~
 \langle N_-\rangle \;=\; \sum_{i,q_i<0} \langle N_i\rangle\;,
 ~~~~
 \langle N_{ch}\rangle \;=\; \sum_{i,q_i\neq 0} \langle N_i\rangle\;,
 \label{pminch}
 }
where $V$ is the system volume and $g_i$ is the degeneracy factor
of particle of species $i$  (a number of spin states). A sum of
the momentum states is transformed into the momentum integral,
which holds in the thermodynamic limit $V\rightarrow \infty$.

The {\it microscopic correlator} in the GCE reads:
\eq{\label{corr-gce}
 \langle \Delta n_{p,i} ~ \Delta n_{k,j}
\rangle
 \;=\;  \upsilon_{p,i}^2\,\delta_{ij}\,\delta_{pk}~,
 }
where $\upsilon_{p,i}^2$ is given by Eq.~(\ref{np-fluc}). This
gives a possibility to calculate the fluctuations of different
observables in the GCE. Note that only particles of the same
species, $i=j$, and on the same level, $p=k$, do correlate in the
GCE. Thus, Eq.~(\ref{corr-gce}) is equivalent to
Eq.~(\ref{np-fluc}): only the Bose and Fermi effects for the
fluctuations of identical particles on the same level are relevant
in the GCE.

In order to include the effect of exact conservation laws, we
introduce the equilibrium probability distribution $W(\Delta
n_{p,i})$ of the deviations of different sets $\{n_{p,i}\}$ of the
occupation numbers from their average value. In the GCE each
$\Delta n_{p,i}$ fluctuates independently according approximately
to the Gauss distribution law for $\Delta n_{p,i}$ with mean
square deviation $\langle \Delta n_{p,i}^2\rangle = v_{p,i}^{ 2}$:
 \eq{
   W_{g.c.e.}(\Delta n_{p,i}) ~\propto ~ \prod_{p,i}
   \exp\left[-~ \frac{\left(\Delta n_{p,i}\right)^2}{2v_{p,i}^{ 2}}
    \right]~.
\label{gauss} }
To justify Eq.~(\ref{gauss}) one can consider (see
Ref.~\cite{step1}) the sum of $n_{p,i}$ in small momentum volume
$(\Delta  p)^3$ with the center at $p$. At fixed $(\Delta p)^3$
and $V\rightarrow \infty$ the average number of particles inside
$(\Delta p)^3$ becomes large. Each particle configuration inside
$(\Delta p)^3$ consists of $(\Delta p)^3 \cdot gV/(2\pi)^{3}\gg 1$
statistically independent terms, each with average value $\langle
n_{p,i}\rangle$ (\ref{np-aver}) and variance $v^{2}_{p,i}$
(\ref{np-fluc}). From the central limit theorem it follows then
that the probability distribution for the fluctuations inside
$(\Delta p)^3$ should be Gaussian. In fact, we always convolve
$n_{p,i}$ with some smooth function  of $p$, so instead of writing
the Gaussian distribution for the sum of $n_{p,i}$ in $(\Delta
p)^3$ we can use it directly for $n_{p,i}$. The next step is to
impose exact conservation laws. The problem is to calculate the
microscopic correlator with three conserved charges, $Q,B,S$, in
the CE, i.e. when global charge conservation laws are imposed on
each microscopic state of the system. The conserved charge, e.g.,
the electric charge $Q$, can be written in the form
$Q\equiv\sum_{p,i}q_i\,n_{p,i}$. An exact conservation law is
introduced as the restriction on the sets  of the occupation
numbers $\{n_{p,i}\}$: only those sets which satisfy the condition
$\Delta Q=\sum_{p,i} q_i\Delta n_{p,i}=0$ can be realized. Then
the distribution (\ref{gauss}) should be modified. This has been
considered before for one conserved charge in the CE
\cite{ce2-fluc} and MCE \cite{mce-fluc}. Now three charge
conservation laws are imposed:
%
\begin{align}\label{gauss-Q}
   W_{c.e.}(\Delta n_{p,i}) ~&\propto ~\prod_{p,i}
   \exp\left[-~ \frac{\left(\Delta n_{p,i}\right)^2}{2v_{p,i}^{ 2}}
      \right] \cdot
    \delta\left(\sum_{p,i} q_i \Delta n_{p,i}^{} \right)\cdot
    \delta\left(\sum_{p,i} b_i \Delta n_{p,i}^{} \right)\cdot
    \delta\left(\sum_{p,i} s_i \Delta n_{p,i}^{} \right)
    \\
 &\propto ~  \int_{-\infty}^{\infty} d \lambda_q d\lambda_b d \lambda_s~\prod_{p,i}
  \exp\left[-~ \frac{\left(\Delta n_{p,i}\right)^2}{2v_{p,i}^{ 2}}
 \;+\; i \lambda_q\; q_i \Delta n_{p,i} \;+\; i \lambda_b\; b_i \Delta n_{p,i}
 \;+\; i \lambda_s\; s_i \Delta n_{p,i} \right]~.
  \nonumber
\end{align}
%
 It is convenient to generalize distribution (\ref{gauss-Q})
 using further the integration along imaginary axis in
 $\lambda$-space.
After completing squares one finds:
%
\eq{& W_{c.e.}(\Delta n_{p,i};~\lambda_q,\lambda_b,\lambda_s)
 ~   \propto \prod_{p,i}
    \exp[~
    - ~ \frac{\left(\Delta n_{p,i} - \lambda_q v_{p,i}^{ 2}q_i
    - \lambda_b v_{p,i}^{ 2}b_i - \lambda_s v_{p,i}^{ 2}s_i\right)^2}
    {2v_{p,i}^{ 2}} \nonumber
 \\
  & \,+\, \frac{\lambda_q^2}{2} v_{p,i}^{ 2}q_i^2
    \,+\, \frac{\lambda_b^2}{2} v_{p,i}^{ 2}b_i^2
    \,+\, \frac{\lambda_s^2}{2} v_{p,i}^{ 2}s_i^2
    \,-\, \lambda_q\lambda_s v_{p,i}^{ 2} q_i\,s_i
    \,-\, \lambda_q\lambda_b v_{p,i}^{ 2} q_i\,b_i
    \,-\, \lambda_b\lambda_s v_{p,i}^{ 2} b_i\,s_i ]\,.\label{W-lambda}
    }
The CE averaging takes the following form:
 \eq{ \label{W-average}
 \langle \;\ldots\; \rangle_{c.e.}
 ~=~\frac{
 \int_{-i\infty}^{i\infty}
     d\lambda_q d\lambda_b d\lambda_s \,
 \int_{-\infty}^{\infty}\prod_{p,i} dn_{p,i}^{}~\ldots ~
    W_{c.e.}(\Delta n_{p,i};~ \lambda_q, \lambda_b, \lambda_s)}
 {\int_{-i\infty}^{i\infty}
     d\lambda_q d\lambda_b d\lambda_s \,
 \int_{-\infty}^{\infty}\prod_{p,i} dn_{p,i}^{}\;
     W_{c.e.}(\Delta n_{p,i};~ \lambda_q, \lambda_b, \lambda_s)}~.
 }
The CE microscopic correlator is as follows (see also Appendix):
 \eq{\label{corr}
& \langle \Delta n_{p,i}  \Delta n_{k,j} \rangle_{c.e.}
 ~=\;  \upsilon_{p,i}^2\,\delta_{ij}\,\delta_{pk}
 \\
& -\;  \frac{\upsilon_{p,i}^2v_{k,j}^2}{|A|}
 \left[q_iq_j M_{qq} + b_ib_j M_{bb} + s_is_j M_{ss} +
 \left(q_is_j + q_js_i\right) M_{qs}
 - \left(q_ib_j + q_jb_i\right) M_{qb} - \left(b_is_j +
b_js_i\right) M_{bs}
 \right]\;, \nonumber
 }
where $|A|$ is the determinant of the matrix,
 \eq{\label{matrix}
 A =
 \begin{pmatrix}
 \Delta (q^2) & \Delta (bq) & \Delta (sq)\\
 \Delta (q b) & \Delta (b^2) & \Delta (sb)\\
 \Delta (q s) & \Delta (b s) & \Delta (s^2)
 \end{pmatrix}\;,
 }
with the following elements, $\;\Delta (q^2)\equiv\sum_{p,i}
q_i^2\upsilon_{p,i}^2\;$, $\;\Delta (qb)\equiv \sum_{p,i}
q_ib_i\upsilon_{p,i}^2\;$, etc.
$M_{ij}$ are the corresponding minors of the matrix $A$, e.g.,
\eq{ M_{qs}~=~det
 \begin{pmatrix}
 \Delta (qb) & \Delta (b^2) \\
 \Delta (qs) & \Delta (bs)
 \end{pmatrix}~.
 }
In the case of conservation of only one (electric) charge, this
reduces to $|A|=\Delta (q^2),\;M_{qq}=1$. To make these formulae
more transparent we write one of the minors explicitely,
%
\eq{
   M_{ss} \;=\; \Delta (q^2)\,\cdot\, \Delta (b^2) \,-\, [\Delta (qb)]^2
   ~ = ~ \left(\sum_{p,i}q^{2}_i~v^{2}_{p,\,i} \right)
   \cdot \left(\sum_{k,j}b^{2}_j~ v^{2}_{k,\,j}  \right)~
   - ~\left(\sum_{p,i} q_i b_i~v^{2}_{p,\,i} \right)^2~,
   }
%
%
The sum, $\sum_{p,i}$~, means integration over momentum $p$, and
summation over hadron-resonance species~$i$.
The microscopic correlator can be also used in the MCE. The exact
energy conservation is imposed with $\delta\left(\Delta
E\right)\equiv \delta\left(\sum_{p,i}\sqrt{m_i^2+p^2}~\Delta
n_{p,i}\right)$. This would lead to additional terms in the r.h.s.
of Eq.~(\ref{corr}) proportional to $\sum_{p,i}(m_i^2+p^2)~
\upsilon_{p,i}^2\;$, $\;\sum_{p,i}\sqrt{m_i^2+p^2}~
q_i~\upsilon_{p,i}^2\;$, etc.

The microscopic correlator (\ref{corr}) can be used  to calculate
correlations and  fluctuations of different physical quantities in
the CE. The first term in the r.h.s. of Eq.~(\ref{corr})
corresponds to the microscopic correlator (\ref{corr-gce}) in the
GCE. The additional terms reflect the (anti)correlations among
different particles, $i\neq j$, and different levels, $p\neq k$,
that appeared due to the global CE charge conservations.  Let us
concentrate on the particle number fluctuations. One can calculate
the correlations in the GCE and CE, respectively,
%
\eq{\label{dNidNj}
 \langle \Delta N_i ~\Delta N_j~\rangle
 ~= \sum_{p,k}~\langle \Delta n_{p,i}~\Delta n_{k,j}\rangle~=~\sum_p~ v_{p,i}^2\;,
~~~~ \langle \Delta N_i ~\Delta N_j~\rangle_{c.e.}
 ~= \sum_{p,k}~\langle \Delta n_{p,i}~\Delta
 n_{k,j}\rangle_{c.e.}\;.
}
The CE scaled variance reads:
\eq{
 \omega^i_{c.e.}~&\equiv~\frac{\langle (\Delta
N_i)^2~\rangle_{c.e.}}{\langle
N_i\rangle_{c.e.}}~=~\omega^i_{g.c.e.}~[~1~\nonumber \\
&-~ \frac{\sum_k ~v_{k,i}^2}{|A|}~
 (q_i^2 M_{qq} + b_i^2 M_{bb} + s_i^2 M_{ss} +
 2q_is_i M_{qs} - 2q_ib_i M_{qb} - 2b_is_i  M_{bs}
 )~]~.\label{omegai-ce}
}
In Eq.~(\ref{omegai-ce}) we used the fact that $\langle
N_i\rangle_{c.e.}$  is equal to $\langle N_i\rangle$
(\ref{Ni-gce}) in the GCE at $V\rightarrow \infty$, and introduced
the scaled variance in the GCE,
\eq{\label{omegai-gce}
\omega^i_{g.c.e.}~\equiv~\frac{\langle (\Delta
N_i)^2~\rangle}{\langle
N_i\rangle}~=~\frac{\sum_p~v^2_{p,i}}{\sum_p~\langle
n_{p,i}\rangle}~.
}
Note that the CE result (\ref{omegai-ce}) is obtained in the
thermodynamic limit, and it does not include a dependence on $V$.
Thus the method can not be used to obtain the finite volume
corrections. A nice feature of the microscopic correlator method
is the fact that particle number fluctuations and correlations in
the CE, being different from those in the GCE, are presented in
terms of quantities calculated within the GCE.

\subsection{Saddle Point Expansion Technique}

In this subsection we discuss the method of treating the CE at
finite volume $V$. Let us for simplicity consider the CE with only
one conserved charge, $Q$, and only one sort of particles with
charges $+1$ and $-1$. The microscopic correlator method
(\ref{omegai-ce},\ref{omegai-gce}) then gives:
\eq{ \label{omegace-pm}
\omega_{c.e.}^{\pm}~=~\frac{\sum_p~v_p^{\pm 2}}{\sum_p\langle
n_p^{\pm} \rangle} \left[1~-~\frac{\sum_k~v_k^{\pm
2}}{\sum_k(v_k^{+ 2}~+~v_k^{- 2})}\right]~,
}
and for zero value of the total net charge, $Q=0$, this reduces to
 \eq{\label{w-ce-Q=0}
 \omega_{c.e.}^{\pm}
 \;=\;\frac{\sum_p~v_p^{\pm 2}}{2\sum_p\langle n_p^{\pm} \rangle}
 \;=\;\frac{1}{2}\,\omega_{g.c.e.}^{\pm}.
 }

Let us start with an example of Boltzmann approximation,
$\gamma\rightarrow 0$. For neutral system, $Q=0$,
one finds in the GCE:
\eq{ \label{gce1}
Z_{g.c.e.}~=~\exp\left(2z\right)~,~~~~
  \langle N_{\pm} \rangle
~=~z~,~~~~
 \langle N^2_{\pm} \rangle
~=~z~+~z^2~,~~~~\omega_{g.c.e.}^{\pm}~=~1~,
}
and in the CE at $V\rightarrow\infty$ \cite{ce-fluc}:
\eq{ \label{ce1}
Z_{c.e.}~=~I_0(2z)~,~~~~
 \langle N_{\pm} \rangle_{c.e.}
~\cong~ z~\left(1~-~\frac{1}{4z}\right)~,~~~~
 \langle N^2_{\pm}
\rangle_{c.e.}
=~ z^2~,
%
}
where $z~
~=~gV(2\pi^2)^{-1}Tm^2K_2(m/T)$ is one particle partition
function.
It then follows:
\eq{
\omega^{\pm}_{c.e.}~\equiv~\frac{\langle N^2_{\pm}
\rangle_{c.e.}~-~\langle N_{\pm} \rangle_{c.e.}^2}{\langle N_{\pm}
\rangle_{c.e.}}~
\cong~
~\frac{1}{2}~\left[1~+~\mathcal{O}(1/V)\right]~\cong~\frac{1}{2}~,
}
which coincides with Eq.~(\ref{w-ce-Q=0}).
%
%
This result has been obtained for the Boltzmann gas. To justify it
for the Bose and Fermi gases we consider now a systematic
saddle-point expansion \cite{ce-b,ce3-fluc,bec} (see also
\cite{mce2-fluc,Hauer}). The CE partition function is defined as
follows \cite{ce-b,ce3-fluc,bec}:
\eq{\label{Zce-phi}
 Z_{c.e.}(Q) \;=\; \sum_{\{n_p^+,n_p^-\}}\exp\left(-\frac{E}{T}\right)
 ~\delta(Q~-~N_+~+~N_-)~=~  \int_0^{2\pi}\frac{d\phi}{2\pi}~ \exp(-iQ\phi)~
  Z_{g.c.e.}(\phi)~,
}
%
with
 \eq{\label{Zgce}
 Z_{g.c.e.}(\phi)
  \;=\; \exp\left(-~\frac{gV}{2\pi^2}\int_0^{\infty}
  \frac{p^2dp}{\gamma}\;
  \left[ \ln\left(1~ -~\gamma \lambda_+ e^{-\varepsilon_p/T+i\phi}\right)
    \;+\;  \ln\left(1~ -~\gamma \lambda_- e^{-\varepsilon_p/T-i\phi}\right)
  \right]\right)~,
}
where $g$ is the degeneracy factor, $\varepsilon_p \equiv
(p^2+m^2)^{1/2}$, and $\gamma =+1$ and $-1$ correspond to Bose and
Fermi statistics, respectively, while the limit $\gamma\rightarrow
0$ gives the Boltzmann approximation. The $\lambda_+$ and
$\lambda_-$ in Eq.~(\ref{Zgce}) are auxiliary parameters that are
set to one in the final formulae. A substitution of $\exp(\pm
i\phi)$ in Eq.~(\ref{Zgce}) by $\exp(\pm\mu/T)$ leads to well
known expression of the GCE partition function, $Z_{g.c.e.}$, with
chemical potential $\mu$ \cite{lan}. In Eq.~(\ref{Zce-phi})
$E=\sum_p\varepsilon_p(n_p^++n_p^-)$ and $N_{\pm}=\sum_p
n_p^{\pm}$. One expands the logarithm in Eq.~(\ref{Zgce}) in the
Taylor series, $-\gamma^{-1}\ln(1-\gamma~
x)=\sum_{n=1}^{\infty}\gamma^{n-1}~x^n/n$. This leads to:
%
\eq{\label{Zce-phi-2}
& Z_{c.e.}(Q) \;=\; \int_0^{2\pi}\frac{d\phi}{2\pi}
 \exp\left(-iQ\phi \;+\;
 \frac{gV}{2\pi^2}\int_0^{\infty}p^2dp\;\sum_{l=1}^{\infty}
 \frac{\gamma^{l-1}}{l}\,e^{-l\varepsilon_p/T}~
 [(\lambda_+~e^{i\phi})^l~+~(\lambda_-~e^{-i\phi})^l]\right)
  \\
 &= \;
 \int_0^{2\pi}\frac{d\phi}{2\pi}
 \exp\left(
 -iQ\phi \;+\;
 \frac{gV}{2\pi^2}\int_0^{\infty}p^2dp\;\sum_{l=1}^{\infty}
 \frac{\gamma^{l-1}}{l}\,e^{-l\varepsilon_p/T}\,
 \left[ \lambda_+^l\sum_{n=0}^{\infty}\frac{(il\varphi)^n}{n!}
 \;+\;  \lambda_-^l\sum_{n=0}^{\infty}\frac{(-il\varphi)^n}{n!}
 \right]\right)~. \nonumber
}
The Boltzmann approximation, $\gamma\rightarrow 0$, corresponds to
only one term, $l=1$, in the sum from Eq.~(\ref{Zce-phi-2}). Using
the following notations,
\eq{\label{cumulant}
 \kappa_n^{\pm} \;&=\; \frac{gV}{2\pi^2}\int_0^{\infty}p^2dp
 \sum_{l=1}^{\infty}~l^{n-1}\gamma^{l-1}~
 e^{-l\;\varepsilon_p/T}~\lambda_{\pm}^l
 \;\equiv\; \sum_{l=1}^{\infty} l^{n-1}z_l^{\pm}~,\\
%
 z_l^{\pm} \;&=\;  \lambda_{\pm}^l\gamma^{l-1}\frac{gV}{2\pi^2}\,
 \frac{Tm^2}{l}\,K_2(l\,m/T)~,
}
where $\kappa_n^{\pm}\propto V$ are the so called cumulants, one
can easily get the following formula:
\eq{\label{Zgce-cumul}
 Z_{c.e.}(Q) \;=\; \exp(\kappa_0^+~+~\kappa_0^-)~\int_0^{2\pi}\frac{d\phi}{2\pi}
 \exp\left[-iQ\phi \;+\;\sum_{n=1}^{\infty}
 \frac{1}{n!}\left(~\kappa_n^+(i\phi)^n+\kappa_n^-(-i\phi)^n~\right)\right]~.
}
%
At $\lambda_{\pm}=\exp(\pm \mu/T)$ the cumulants $\kappa_l^{\pm}$
give the GCE values:
 %
 \eq{\label{cumulants}
 \kappa_1^{\pm} = \sum_p\langle n_p^{\pm}\rangle = \langle
 N_{\pm}\rangle\;,~~~~
 \kappa_2^{\pm}
 = \sum_p v_p^{\pm\,2} \equiv \langle(\Delta N_{\pm})^2\rangle~,~~~~
 \kappa_3^{\pm}= \sum_p\langle (n_p^{\pm} -  \langle n_p^{\pm}\rangle)^3\rangle
 \;,
\ldots
}
The average values and fluctuations in the CE can be obtained as
the following:
%
\eq{\label{Nce}
 \langle N_{\pm}\rangle_{c.e.}
 \;\equiv\; \left[\frac{1}{Z_{c.e.}}
 \lambda_{\pm}\frac{\partial Z_{c.e.}}{\partial
 \lambda_{\pm}}\right]_{\lambda_{\pm}=1}~,~~~~
  \langle N_{\pm}^2\rangle_{c.e.}
  \;\equiv\; \left[\frac{1}{Z_{c.e.}}\lambda_{\pm}\frac{\partial}{\partial \lambda_{\pm}}
 \left(\lambda_{\pm}\frac{\partial Z_{c.e.}}{\partial \lambda_{\pm}}
 \right)\right]_{\lambda_{\pm}=1}~.
 }
To calculate (\ref{Nce}) one needs to estimate the following
integrals,
 \eq{\label{I-Q}
 I(\widetilde{Q}) \;=\; \int_0^{2\pi} d\phi~ \exp\left(-i\widetilde{Q}\phi
 -\kappa_2\phi^2
 \;+\; \frac{\kappa_4}{12}\phi^4 \;-\; \frac{\kappa_6}{360}\phi^6 \;+\; \ldots
 \right)~.
 }
At $V\rightarrow\infty$ an integrand in (\ref{I-Q}) has a strong
maximum at $\varphi=0$, which leads to the result:
 \eq{\label{Z-Q}
 I(\widetilde{Q})\;\propto \; \left(1\;-\;\frac{\widetilde{Q}^2}
 {4\kappa_2} \;+\; \frac{1}{16}\frac{\kappa_4}{\kappa_2^2}
  \right) \;\equiv\; Z_{\widetilde{Q}}\;.
  }
The Eq.~(\ref{Nce}) then leads to
\cite{ce3-fluc,bec}:
 \eq{ \label{Nce1}
 \langle N_{\pm}\rangle_{c.e.}
 &\;=\; \sum_{n=1}^{\infty}z_n\frac{Z_{Q\mp n}}{Z_{Q}}~,
 \\
 \langle N_{\pm}^2\rangle_{c.e.}
 & \;=\; \sum_{n=1}^{\infty}z_nn\frac{Z_{Q\mp n}}{Z_{Q}}\;+\;
 \sum_{l=1}^{\infty}\sum_{n=1}^{\infty}z_lz_n\frac{Z_{Q\mp
 (l+n)}}{Z_{Q}}~.\label{N2ce1}
 }
At $Q=0$, Eqs.~(\ref{Nce1}) and (\ref{N2ce1}) read:
 \eq{
 \label{Nce2}
 \langle N_{\pm}\rangle_{c.e.}
 &\;=\; \sum_{n=1}^{\infty} z_n\left(1-\frac{n^2}{4\kappa_2}
 \;+\;\mathcal{O}(V^{-2})\right) \;=\;  \kappa_1\left(1-\frac{\kappa_3}{4\kappa_1\kappa_2}
 \;+\;\mathcal{O}(V^{-2})\right)~,
 \\
 \langle N_{\pm}^2\rangle_{c.e.}
 &\;=\; \sum_{n=1}^{\infty} z_nn\left(1-\frac{n^2}{4\kappa_2} +\mathcal{O}(V^{-2})\right)
 \;+\; \sum_{l=1}^{\infty}\sum_{n=1}^{\infty}
 z_lz_n\left(1-\frac{(l+n)^2}{4\kappa_2} +\mathcal{O}(V^{-2})\right)
 \nonumber \\
 &\;=\; \kappa_1^2\left(1-\frac{\kappa_3}{2\kappa_1\kappa_2} \;+\;
 \frac{\kappa_2}{2\kappa_1^2}\;+\;\mathcal{O}(V^{-2})\right)~.
 \label{N2ce2}
 }
%
%
The scaled variance equals:
 \eq{
 \omega^{\pm}_{c.e.} \;=\;\frac{\kappa_2}{2\kappa_1}
 \left[1\;+\;\mathcal{O}(V^{-1})\right]
 \;=\;
 \frac{1}{2}\;\omega^{\pm}_{g.c.e.}\left[1\;+\;\mathcal{O}(V^{-1})\right]~,
 }
which coincides with Eq.~(\ref{w-ce-Q=0}) in the large volume
limit $V\rightarrow\infty$.

One again observes that the global conservation laws lead to the
correction to average particle numbers, $\langle
N_{\pm}\rangle_{c.e.}= \langle N_{\pm}\rangle
[1-\mathcal{O}(1/V)]$. It equals $\kappa_3/(4\kappa_1\kappa_2)$
and leads to additional terms to $\langle N_{\pm}\rangle^2_{c.e.}$
and $\langle N_{\pm}^2\rangle_{c.e.}$ proportional to $V$. These
terms, however, are cancelled out in the variances, and
Eq.~(\ref{w-ce-Q=0}) obtained from the microscopic correlator
remains valid. This gives justification of the microscopic
correlator approach, which assumes the equality $\langle
n_{p,i}\rangle_{c.e.}=\langle n_{p,i}\rangle$  in the
thermodynamic limit $V\rightarrow\infty$.

\section{Effect of Resonance Decays}\label{sec-res}

\subsection{Generating Function}
Resonance decay has a probabilistic character. This itself causes
the particle number fluctuations in the final state. The average
number of final particles from resonance decays, and all higher
moments including particle correlations can be found from the
following generating function:
 \eq{\label{G}
 G
 \;\equiv \; \prod_R \left(\sum_{r}b_r^R~\prod_{i}\lambda_i^{n_{i,r}^R}
\right)^{N_R}\,,
 }
where $b^R_r$ is the branching ratio of the $r$-th branch,
$n_{i,r}^R$ is the number of $i$-th particles produced in that
decay mode, and $r$ runs over all branches  with the requirement
$\sum_{r} b_r^R=1$. Note that different branches are defined in a
way that final states with only stable (with respect to strong and
electromagnetic decays) hadrons are counted. The $\lambda_i$ in
Eq.~(\ref{G}) are auxiliary parameters that are set to one in the
final formulae.
The averages from resonance decays can be found as the following:
 \eq{\label{N-i}
 \overline{N_i}~&
 \equiv~ \sum_R \langle N_i\rangle_R \;=\; \lambda_i\frac{\partial}{\partial
 \lambda_i}~G ~=~  \sum_R N_R~ \sum_r b_r^R
n_{i,r}^R
 \;\equiv\;  \sum_R N_R~ \langle n_{i}\rangle_R~,\\
 \overline{N_i~N_j}~& \equiv~\sum_R\langle N_i~N_j\rangle_R
 \;=\;  \lambda_i\frac{\partial}{\partial
 \lambda_i} \left(\lambda_j\frac{\partial}{\partial
 \lambda_j}~G\right)\nonumber \\
 &=\;
 \sum_R \left[~ N_R~(N_R-1)\; \langle n_i\rangle_R\, \langle n_j\rangle_R
 \;+\;  N_R ~ \langle n_i~n_j\rangle_R~ \right]~, \label{NiNj}
}
where $\langle n_i~n_j\rangle_R \equiv \sum_r b_r^R n_{i,r}^R
n_{j,r}^R$\;.
%
%
%
The averaging, $\langle \cdots\rangle_R$, in Eq.~(\ref{N-i}) means
the averaging over resonance decays. The formula (\ref{G})
originates from the fact that the normalized probability
distribution, $P(N_R^r)$, for the decay of $N_R$ resonances  is
the following:
 \eq{\label{multinom}
 P(N_R^r) \;=\; N_R!\;
    \prod_r \frac{ (b_r^R)^{N^r_R}}{ N^r_R!}\;
     \delta\left(\sum_r N^r_R\;-\;N_R\right)\;,
 }
where $N^r_R$ correspond to the numbers of $R$-th resonances
decaying via $r$-th branch.

The scaled variance $\omega_R^{i*}$ due to decays of $R$-th
resonances reads:
 \eq{\label{omegai-R}
 \omega_R^{i*}
 \;\equiv\; \frac{\langle N_{i}^2\rangle_R \;-\; \langle N_{i}\rangle_R^2}
            {\langle N_{i}\rangle_R}
 \;=\; \frac{\langle n_i^{2}\rangle_R \;-\; \langle n_i\rangle_R^2}
       {\langle n_i\rangle_R}
 ~\equiv~
       \frac{\sum_r b_r^R~ (n_{i,r}^{R})^2 \;-
       \; (\sum_r b_r^R ~n_{i,r}^R)^2}
       {\sum_r b_r^R~ n_{i,r}^R~}~.
 }
To illustrate Eq.~(\ref{omegai-R}) some examples are appropriate.
%
%
It follows from Eq.~(\ref{omegai-R}) that $\omega_R^{i*}=0$ if
$n^R_{i,r}$ were the same in all decay channels. The
$\omega_R^{i*}$ also vanishes if there was only one decay channel,
i.e. $b_1^R=1$.
%
%
Let there be an arbitrary number of $x$-th type decay channels
with $n^R_{i,x}=1$ and $y$-th type ones with $n^R_{i,y}=0$. From
Eq.~(\ref{omegai-R}) one finds $\omega_R^{i*}=1-b^R_x>0$, where
$b^R_x$ is the total probability of $x$-th type decay channels. If
$n^R_{i,x}=2$ and $n^R_{i,y}=0$, then $\omega_R^{i*}=2(1-b^R_x)$.
%
%
%
In general, Eq.~(\ref{omegai-R}) tells that resonance decays
generate fluctuations of $i$-th hadron multiplicity if $n^R_{i,r}$
are different in different decay channels. If $n^R_{i,r}$ is
larger than 1 in some of these channels, the fluctuations become
stronger.

The Eqs.~(\ref{N-i},\ref{NiNj}) assume some fixed values of $N_R$.
In a real situation, $N_R$ fluctuate, and this is an additional
source of the particle number fluctuations. One finds:
 \eq{
 \omega_R^i~\equiv~\frac{\langle\langle N_i^2\rangle_R\rangle_T~-~
\langle\langle N_i\rangle_R\rangle^2_T}{\langle\langle
N_i\rangle_R\rangle_T}
 \;=\;\omega^{i*}_R \;+\; \langle
 n_i\rangle_R\,\omega_{R}\;,
 }
where resonances act as sources of particles, similar to the  so
called independent source model \cite{fluc7}, and the scaled
variance,
 \eq{
\omega_{R} \;\equiv\; \frac{\langle N_R^2\rangle_{T} \;-\; \langle
N_R\rangle^2_T}{\langle N_R\rangle_{T}}~,
 }
corresponds to the thermal (GCE or CE) fluctuation of the number
of resonances.

\subsection{Grand Canonical Ensemble}

The average number of $i$-particles in the presence of primary
particles $N_i^*$ and different resonance types $R$ is the
following:
 \eq{\label{<N>}
 \langle N_i\rangle
 \;=\; \langle N_i^*\rangle + \sum_R \langle N_R\rangle \sum_r b_r^R n_{i,r}^R
 \;\equiv\; \langle N_i^*\rangle + \sum_R \langle N_R\rangle \langle n_{i}\rangle_R
 }
The summation $\sum_R$ runs over all types of resonances. The
$\langle\ldots\rangle$ and $\langle\ldots\rangle_R$ correspond to
the GCE averaging, and that over resonance decay channels.

From Eqs.~(\ref{G}, \ref{N-i}) one finds the GCE correlators
\cite{Jeon-Koch}:
%
%
 \eq{\label{corr-GCE}
  \langle \Delta N_i\,\Delta N_j\rangle~=~
  \langle\Delta N_i^* \Delta N_j^*\rangle
  \;+\; \sum_R \left[ \langle\Delta N_R^2\rangle\;
  \langle n_{i}\rangle_R\;\langle n_{j}\rangle_R
  \;+\; \langle N_R\rangle\; \langle \Delta n_{i}\Delta n_{j}\rangle_R
  \right]~.
 }
The terms proportional to $\langle N_R\rangle \langle N_{R'}\rangle$
and $\langle N_i^*\rangle \langle N_R\rangle$  cancel each other in
the GCE calculations of the correlator (\ref{corr-GCE}).

\subsection{Canonical Ensemble}

All primary particles and resonances become to correlate in the
presence of exact charge conservation laws. Thus for the CE
correlators
we obtain a new result:
 \eq{\label{corr-CE}
 & \langle \Delta N_i\,\Delta N_j\rangle_{c.e.}
 \;=\; \langle\Delta N_i^* \Delta N_j^*\rangle_{c.e.}
  \;+\; \sum_R \langle N_R\rangle\; \langle \Delta n_{i}\; \Delta  n_{j}\rangle_R
 \;+\; \sum_R \langle\Delta N_i^*\; \Delta N_R\rangle_{c.e.}\; \langle n_{j}\rangle_R
  \; \nonumber
 \\
 &+\; \sum_R  \langle\Delta N_j^*\;\Delta N_R\rangle_{c.e.}\; \langle n_{i}\rangle_R
  \;+\; \sum_{R, R'} \langle\Delta N_R\;\Delta N_{R'}\rangle_{c.e.}
  \; \langle n_{i}\rangle_R\;
       \langle n_{j}\rangle_{R^{'}}\;.
 }
Additional terms in Eq.~(\ref{corr-CE}) compared to
Eq.~(\ref{corr-GCE}) are due to the correlations induced by exact
charge conservations in the CE. The Eq.~(\ref{corr-CE}) remains
valid in the MCE too with $\langle \ldots \rangle_{c.e.}$ replaced
by $\langle \ldots \rangle_{m.c.e.}$.

\section{scaled variances along the chemical freeze-out line}\label{sec-HG}
In this section we present calculations of the CE and GCE
fluctuations along the chemical freeze-out line in central Pb+Pb
(Au+Au) for both primordial and final state distributions.

At chemical freeze-out the hadronic gas is usually described by
the following parameters:  temperature $T$, chemical potentials
($\mu_B$, $\mu_S$, $\mu_Q$), and strangeness suppression factor
$\gamma_S$ to account for a undersaturation of the strange sector.
The GCE has proven to be sufficient for thermal model analysis of
mean multiplicities in central Pb-Pb and Au-Au collisions at most
colliding energies. Only at lower energies, where only a few
strange particles are produced, the CE effects of an exact
strangeness conservation become visible. This leads to the CE
suppression of yields of strange particles  when compared to the
GCE. In the energy range discussed below this is only the case for
the SIS data point. On the other hand,  for multiplicity
fluctuations the exact conservation laws are important for all
colliding energies.

Thermal model analysis has provided a systematic evolution of the
parameter set with beam energy and size of colliding system and
allows for phenomenological parametrization, giving the thermal
model almost predictive qualities. A recent discussion of system
size and  energy dependence of freeze-out parameters and
comparison of freeze-out criteria can be found in Refs.~
\cite{FOP,FOC}.

There are several programs designed for the statistical analysis
of particle production in relativistic heavy-ion collisions, see
e.g., SHARE \cite{Share} and THERMUS \cite{Thermus}. In this paper
an extended version of the THERMUS thermal model framework
\cite{Thermus} is used. With increasing colliding energy, the
temperature increases and more energy for particle production
becomes available. This is accompanied by a drop in $\mu_B$, which
can be parameterized by the following function \cite{FOC}:
\begin{equation}
\mu_B \left( \sqrt{s_{NN}} \right) = \frac{1.308~\mbox{GeV}}{1+
0.273~ \sqrt{s_{NN}}}~, \label{muB}
\end{equation}
where the c.m. nucleon-nucleon collision energy, $\sqrt{s_{NN}}$,
is taken in GeV units in Eq.~(\ref{muB}).

The electrical chemical potential $\mu_Q$ can be further adjusted
to give the charge to baryon ratio of heavy nuclei, $Q/B \approx
0.4$. Strange chemical potential $\mu_S$ is constrained  by
requiring the system to be net strangeness free, $S=0$. Finally
the temperature is chosen to match a condition, $\langle E
\rangle/\langle N \rangle \approx 1~$GeV \cite{Cl-Red}, for energy
per hadron.  In order to remove the remaining free parameter,
$\gamma_S$, we use the following parametrization \cite{FOP}:
\begin{equation}
\gamma_S~ =~ 1 - 0.396~ \exp \left(  - ~\frac{1.23~ T}{\mu_B}
\right)~.
\end{equation}
Numerical fitting functions allow to meet all the above criteria
simultaneously and thus to choose a parameter set, $(T,\mu_B)$,
for each given collision energy. The corresponding chemical
freeze-out line in the $T-\mu_B$ plane is shown in Fig.~1. There
is obviously some degree of freedom as to choose a particular
parametrization for some parameter or value of the average energy
per particle. This particular choice is in good agreement with
thermal model fits done in Ref.~\cite{FOP}. The center of mass
nucleon-nucleon energies, $\sqrt{s_{NN}}$, quoted in Table~I
correspond to beam energies at SIS (2~AGeV), AGS (11.6~AGeV), SPS
(20, $30$, $40$, $80$, and $158$~AGeV), and two top colliding
energies at RHIC ($\sqrt{s_{NN}}=130$~GeV and $200$~GeV).
\begin{table}[h!]
\begin{center}
\begin{tabular}{||c||c|c|c|c||}\hline
$\sqrt{s_{NN}} \left[GeV \right]$ & $T[MeV]$ & $\mu_B[MeV]$
&$\gamma_S$ & $\rho_B \left[ fm^{-3} \right]$\\ [0.5ex] \hline
\hline
$2.32$ & 64.9 & 800.8 & 0.642 & 0.061\\
$4.86$ & 118.5 & 562.2 & 0.694 & 0.111\\
$6.27$ & 130.7 & 482.4 & 0.716 & 0.117\\
$7.62$ & 138.3 & 424.6 & 0.735 & 0.117\\
$8.77$ & 142.9 & 385.4 & 0.749 & 0.115\\
$12.3$ & 151.5 & 300.1 & 0.787 & 0.104\\
$17.3$ & 157.0 & 228.6 & 0.830 & 0.088\\
$130$ & 163.6 & 35.8 & 0.999 & 0.016\\
$200$ & 163.7 & 23.5 & 1 & 0.010\\
\hline
\end{tabular}
\label{ParameterTable} \caption{Chemical freeze-out parameters for
central Pb+Pb (Au+Au) collisions.}
\end{center}

\end{table}

\begin{figure}[ht!]
\begin{center}
\epsfig{file=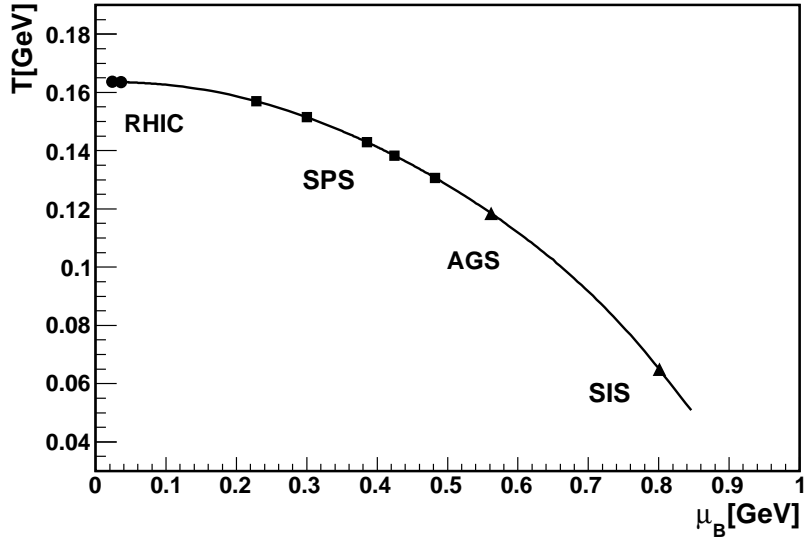,width=12cm} \caption{Chemical freeze-out
curve for central Pb+Pb (Au+Au) collisions.} \label{TmuB}
\end{center}
\end{figure}

\begin{figure}[ht!]
\begin{center}
 \epsfig{file=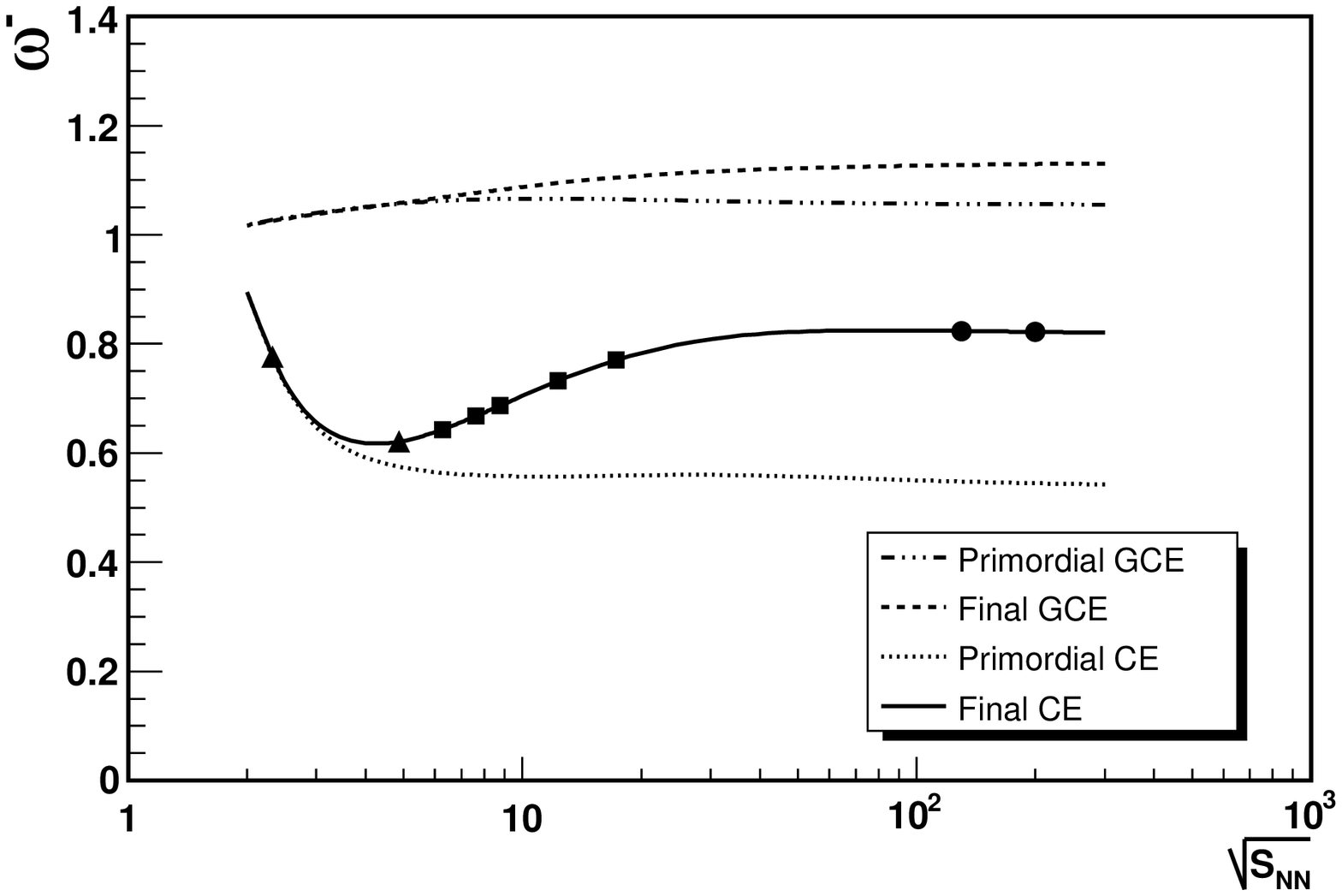,width=12cm}
 \caption{The scaled variances for negatively charged
 particles, $\omega^-$, along the chemical freeze-out line for
central Pb+Pb (Au+Au) collisions (see Fig.~\ref{TmuB}).
 Different lines present primordial and final GCE and CE results.} \label{omega_m}
\end{center}

\end{figure}

\begin{figure}[ht!]
\begin{center} \epsfig{file=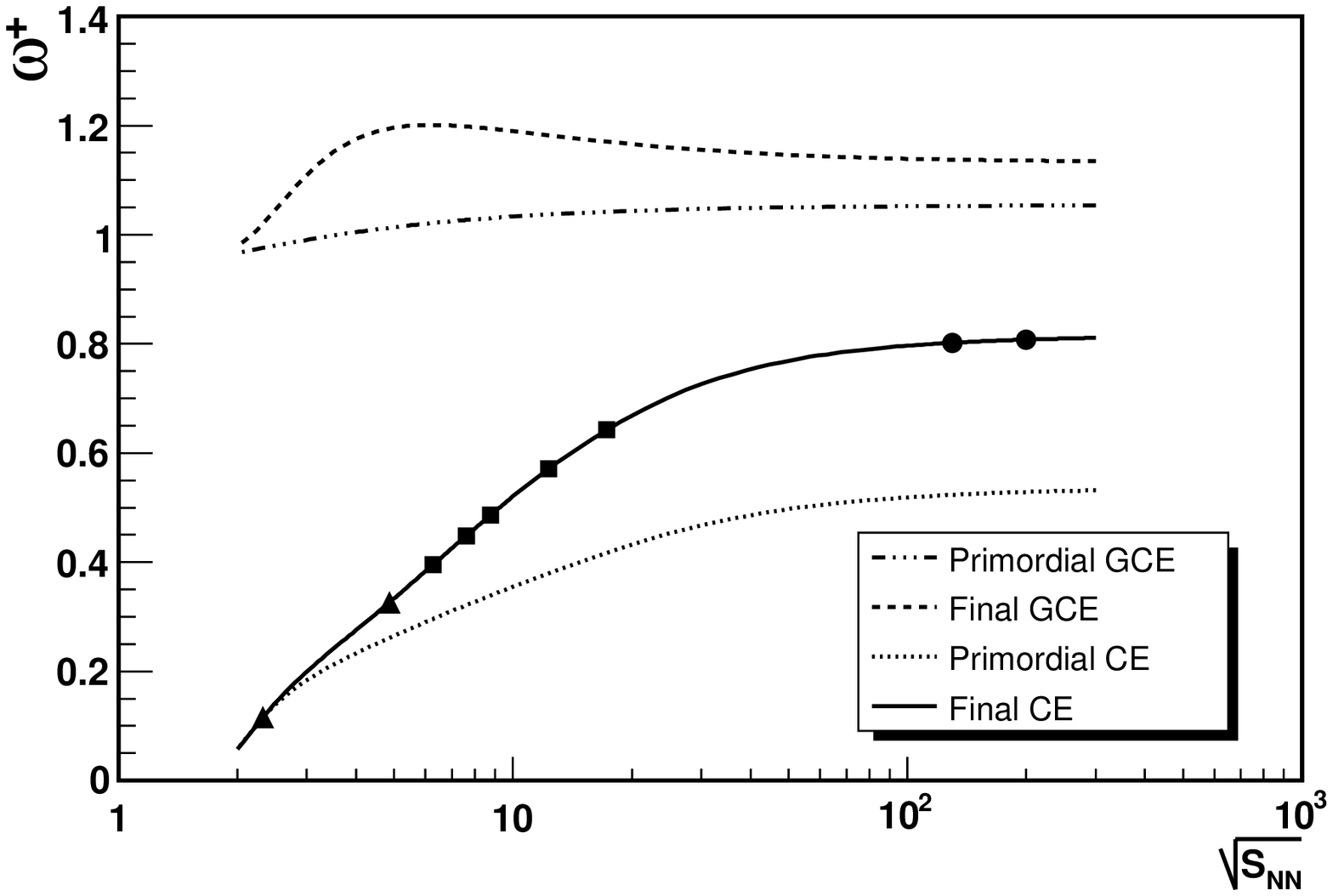,width=12cm}
 \caption{The scaled variances for positively charged
 particles, $\omega^+$, along the chemical freeze-out line  for
central Pb+Pb (Au+Au) collisions (see Fig.~\ref{TmuB}).
 Different lines present primordial and final GCE and CE results.} \label{omega_p}
\end{center}

\end{figure}

\begin{figure}[ht!]
\begin{center} \epsfig{file=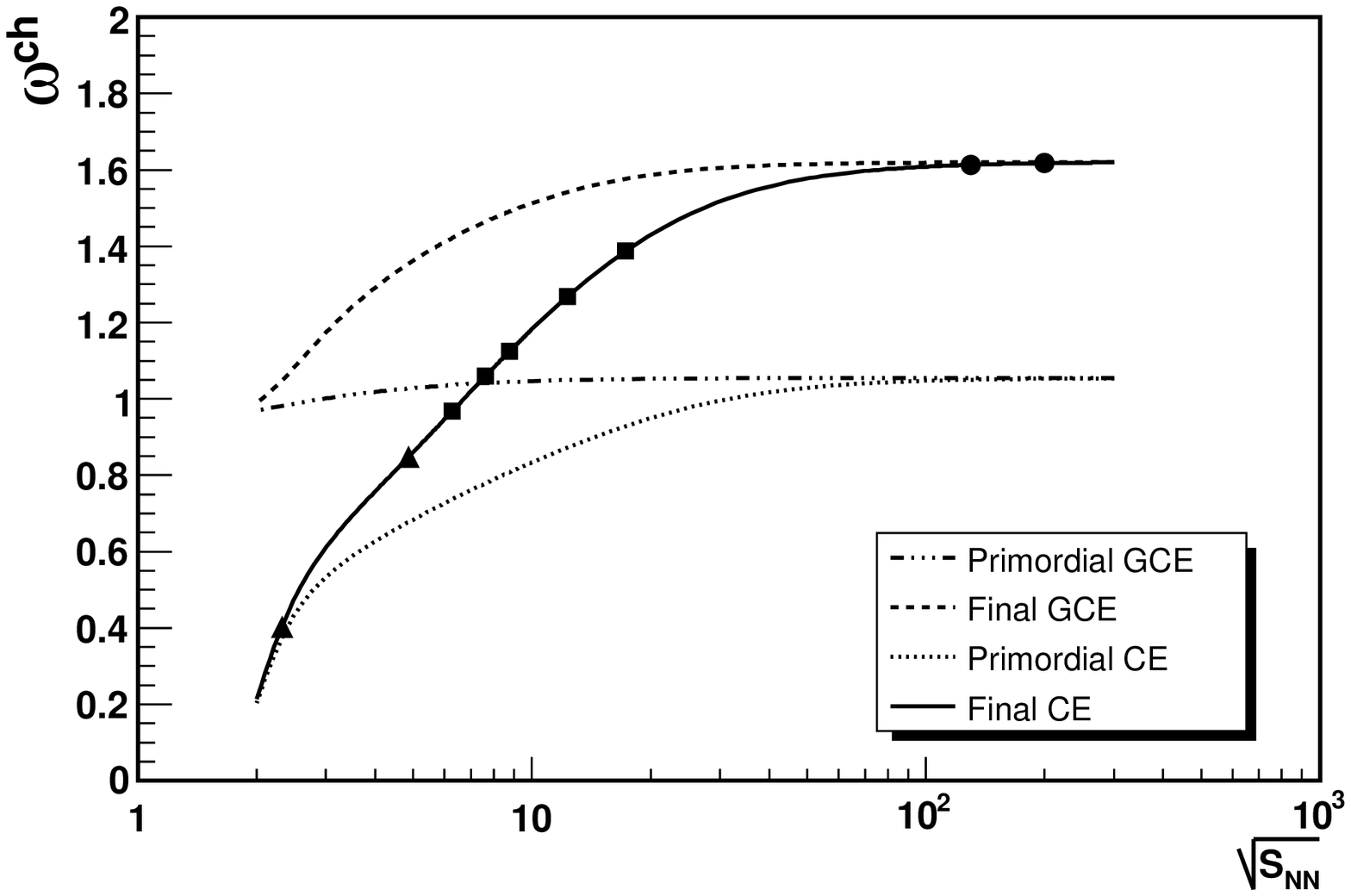,width=12cm}
 \caption{The scaled variances for all  charged
 particles, $\omega^{ch}$, along the chemical freeze-out line  for
central Pb+Pb (Au+Au) collisions (see Fig.~\ref{TmuB}).
 Different lines present primordial and final GCE and CE results.} \label{omega_ch}
\end{center}
\end{figure}

Once a suitable parameter set is determined, mean occupation
numbers and fluctuations can be calculated using
Eqs.~(\ref{np-aver}) and ~(\ref{np-fluc}). The scaled variances of
negative, positive, and all charged particles read:
\eq{\label{omega-all}
 \omega^- ~=~ \frac{\langle \left( \Delta N_- \right)^2
\rangle}{\langle N_-
  \rangle}~,
~~~~\omega^+~ =~ \frac{\langle \left( \Delta N_+ \right)^2
\rangle}{\langle N_+
  \rangle}~,~~~~
\omega^{ch} ~=~ \frac{\langle \left( \Delta N_{ch} \right)^2
\rangle}{\langle N_{ch}
  \rangle}~,
}
where
\eq{
\langle \left( \Delta N_- \right)^2 \rangle~& =~
\sum_{i,j;~q_i<0,q_j<0} \langle \Delta N_i \Delta N_j
\rangle~,~~~~ \langle \left( \Delta N_+ \right)^2 \rangle~ =~
\sum_{i,j;~q_i>0,q_j>0} \langle \Delta N_i \Delta N_j \rangle~, \label{DNpm}\\
\langle \left( \Delta N_{ch} \right)^2 \rangle~& =~
\sum_{i,j;~q_i\neq 0,q_j\neq 0} \langle \Delta N_i \Delta N_j
\rangle~. \label{DNch}
}
For the primordial hadrons, mean multiplicities, $\langle
N_i\rangle$, in Eq.~(\ref{omega-all}) are given by
Eq.(\ref{Ni-gce}), and correlators, $\langle \Delta N_i \Delta N_j
\rangle$, in Eqs.~(\ref{DNpm},\ref{DNch}) are either the GCE or CE
correlators from Eq.~(\ref{dNidNj}).  For final state mean
multiplicities, Eq.~(\ref{<N>}) is used, and correlators are
calculated with Eq.~(\ref{corr-GCE}) in the GCE or
Eq.~(\ref{corr-CE}) in the CE, respectively. For final hadrons the
summation needs to be extended to all stable particles with
corresponding charges and all unstable resonances which have these
charged particles in their decay channels. Figures \ref{omega_m},
\ref{omega_p}, and \ref{omega_ch} show scaled variances for
negatively charged particles, $\omega^-$, positively charged
particles, $\omega^+$, and all charged particles, $\omega^{ch}$,
respectively, as functions of $\sqrt{s_{NN}}$. Four cases are
considered, namely, primordial and final state particles in both
the GCE and CE.
The relevant primordial and  final state values for various
colliding energies are summarized in Tables \ref{OmegaTablePrim}
and \ref{OmegaTableFinal}, respectively.

\begin{table}[h!]
\begin{center}
\begin{tabular}{||c||c|c||c|c||c|c||}\hline
 & \multicolumn{2}{c||}{ $\omega^{ch}$} & \multicolumn{2}{c||}{
  $\omega^{+}$} & \multicolumn{2}{c||} { $\omega^{-}$} \\ [0.5ex]
\hline $\sqrt{s_{NN}} \left[GeV \right]$ & GCE & CE & GCE & CE&
GCE & CE \\ [0.5ex] \hline\hline
$2.32$ & 0.982 & 0.373 & 0.976 & 0.115& 1.027 & 0.775 \\
$4.86$ & 1.027 & 0.677 & 1.013 & 0.261& 1.059 & 0.575 \\
$6.27$ & 1.036 & 0.737 & 1.021 & 0.296& 1.062 & 0.564 \\
$7.62$ & 1.041 & 0.779 & 1.027 & 0.321 & 1.065 & 0.560 \\
$8.77$ & 1.044 & 0.808 & 1.030 & 0.339 & 1.066 & 0.558 \\
$12.3$ & 1.049 & 0.872 & 1.037 & 0.380 & 1.066 & 0.557 \\
$17.3$ & 1.052 & 0.929 & 1.042 & 0.418& 1.065 & 0.559 \\
$130$ & 1.054 & 1.050 & 1.053 & 0.523& 1.056 & 0.548 \\
$200$ & 1.055 & 1.053 & 1.053 & 0.529& 1.056 & 0.545 \\
\hline
\end{tabular}
\caption{Primordial scaled variances in the GCE and CE for central
Pb+Pb (Au+Au) collisions.} \label{OmegaTablePrim}
\end{center}
\end{table}

\begin{table}[h!]
\begin{center}
\begin{tabular}{||c||c|c||c|c||c|c||}\hline
 & \multicolumn{2}{c||}{ $\omega^{ch}$} & \multicolumn{2}{c||}{
  $\omega^{+}$} & \multicolumn{2}{c||} { $\omega^{-}$} \\ [0.5ex]
\hline $\sqrt{s_{NN}} \left[GeV \right]$ & GCE & CE & GCE & CE&
GCE & CE \\ [0.5ex] \hline\hline
$2.32$ & 1.048 & 0.403 & 1.020 & 0.116& 1.025 & 0.777 \\
$4.86$ & 1.354 & 0.848 & 1.195 & 0.327& 1.058 & 0.621 \\
$6.27$ & 1.421 & 0.967 & 1.201 & 0.395& 1.068 & 0.643 \\
$7.62$ & 1.464 & 1.059 & 1.198 & 0.449 & 1.076 & 0.668 \\
$8.77$ & 1.491 & 1.124 & 1.194 & 0.486 & 1.082 & 0.687 \\
$12.3$ & 1.542 & 1.268 & 1.182 & 0.571 & 1.095 & 0.732 \\
$17.3$ & 1.576 & 1.387 & 1.171 & 0.643 & 1.105 & 0.770 \\
$130$ & 1.619 & 1.613 & 1.138 & 0.802& 1.128 & 0.824 \\
$200$ & 1.620 & 1.617 & 1.136 & 0.808& 1.130 & 0.822 \\
\hline
\end{tabular}
\caption{Final state scaled variances in the GCE and CE for
central Pb+Pb (Au+Au) collisions.} \label{OmegaTableFinal}
\end{center}
\end{table}

The column $\rho_B$ in Table~I allows for a comparison with
previously reported values of primordial scaled variances
\cite{bec} (a good agreement is found). The standard THERMUS
particle table includes all strange and light flavored particles
and resonances up to about $2.6$~GeV. Only strong and
electromagnetic decays are considered, weakly decaying channels
are omitted. It should be mentioned that, in particular, heavy
resonances do not always have well established decay channels,
thus there are always some ambiguities in the implementation of
resonance decays in respective thermal model codes. Details about
the THERMUS decay convention can be found in Ref.~\cite{Thermus}.
The quoted values of the scaled variances are valid in the
thermodynamic limit and assume that all charge carriers are
detected.
For high $\mu_B$ (low collision energies) the multiplicity of
positively charged particles, $N_+$, is enhanced in a comparison
with $N_-$, while the fluctuations $\omega^+$ are suppressed in a
comparison with $\omega^-$. At vanishing net charge density (high
collision energies), $\omega^+$ and $\omega^-$ have the same
asymptotic values. The scaled variance for all charged particles
$\omega^{ch}$ has the same value in the GCE and CE for a neutral
system, for both primordial and final state. The effect of
resonance decays remains small at low collision energies (i.e.
small temperatures), while becoming sizeable even at the lowest
SPS energy.

Some important qualitative effects are seen in
Figs.~\ref{omega_m}, \ref{omega_p}, and \ref{omega_ch}. The effect
of Bose and Fermi statistics can be seen in primordial values in
the GCE. At low temperatures Fermi statistics dominate,
$\omega^{+}_{g.c.e.}, \omega^{ch}_{g.c.e.}<1$, while at high
temperature (low $\mu_B$) Bose statistics dominate,
$\omega_{g.c.e.}^{\pm}, \omega_{g.c.e.}^{ch}>1$. At the chemical
freeze-out line, $\omega_{g.c.e.}^-$ is always slightly larger
than 1, as $\pi^-$ is the dominant negative particle at low
temperature too. A bump at small collision energies in
$\omega^+_{g.c.e.}$ for final particles is due to the
$\Delta^{++}$ decay into 2 positively charged hadrons, $p+\pi^+$.
This single resonance contribution dominates at small collision
energies (temperatures), but becomes relatively unimportant at
high collision energies. A minimum in $\omega_{c.e.}^{-}$ for
final particles is seen in Fig.~\ref{omega_m}. This happens as a
result of the following effects. Since the number of negative
particles is relatively small, $\langle N_-\rangle \ll \langle
N_+\rangle$, at low collision energies, the CE suppression effects
are also small. Low collision energies correspond to small
temperatures of the hadron-resonance system, and the resonance
decay effects are small too. With increasing $\sqrt{s_{NN}}$, the
CE effects increase and this makes $\omega^-_{c.e}$ smaller, but
resonance decay effects increase too and they work in an opposite
direction making $\omega^-_{c.e}$ larger.
A combination of these two effects, CE suppression and resonance
enhancement, leads to a minimum structure of $\omega^-_{c.e}$ seen
in Fig.~\ref{omega_m}.

The results for scaled variances presented in Figs.~2--4 and
Tables II, III correspond to an ideal situation when all final
hadrons are accepted by the detector.  To compare our calculations
to experimentally obtained values of $\omega$ the acceptance and
resolution need to be taken into account. Observing only a
fraction $q$ of final state particles dilutes the effect of global
charge conservation. Even though the primordial particles at the
chemical freeze-out line are only weakly correlated in the
momentum space,  this is no longer valid for final state particles
as the decay products of resonances are not re-thermalized.
Neglecting the momentum correlations due to resonance decays (this
is approximately valid for $\omega^+$ and $\omega^-$, and much
worse for $\omega^{ch}$) the following approximation for the
scaled variances of experimentally accepted particles can be used
(see e.g., \cite{ce-fluc,fluc7}),
\begin{equation}\label{acc}
\omega_{acc}~ =~ 1~ -~ q ~+ ~q ~\omega_{4 \pi}~,
\end{equation}
where $\omega_{4 \pi}$ refers to an ideal detector with full
$4\pi$-acceptance. In the limit of a very `bad' (or `small')
detector, $q\rightarrow 0$, all scaled variances approach linearly
to 1, i.e., this would lead to the Piossonian distributions for
detected particles.

\section{Summary}\label{sec-summary}
The  multiplicity fluctuations of hadrons in relativistic
nucleus-nucleus collisions have been considered in the statistical
model within the canonical ensemble formulation. The microscopic
correlator method, previously used for one conserved charge, has
been extended to include three conserved charges -- baryon number,
electric charge, and strangeness. The {\it analytical formulae}
for the resonance decay contributions to the correlations and
fluctuations have been found. Using the full hadron-resonance
spectrum we have calculated the scaled variances of negative,
positive and all charged particles for primordial and final
hadrons at the chemical freeze-out in central Pb+Pb (Au+Au)
collisions for different collision energies from SIS and AGS to
SPS and RHIC. Both the CE and resonance decay effects for the
multiplicity fluctuations have been discussed. A comparison with
the NA49 data in Pb+Pb collisions at the SPS energies can be done
for the sample of most central events with the number of
projectile participants being close to its maximal value,
$N_P^{proj}\approx A$, to avoid the fluctuations of the number of
nucleon participants (see discussion in Ref.~\cite{MGMG,KGB}).
These NA49 data will be available soon \cite{GL}. The predictions
of the statistical model within the CE formulations can be done
for $\omega^-$ and $\omega^+$. In this case the experimental
acceptance can be approximately introduced by a simple procedure
based on Eq.~(\ref{acc}). We find a qualitative difference between
the CE results, $\omega^{\pm}_{c.e.}<1$, and the GCE ones,
$\omega_{g.c.e.}^{\pm}> 1$, for the accepted particles.

\begin{acknowledgments}
We would like to thank  F. Becattini, E.L.~Bratkovskaya,
A.I.~Bugrij, M. Ga\'zdzicki, A.P.~Kostyuk, B.~Lungwitz,
I.N.~Mishustin, St.~Mr\'owczy\'nski, L.M.~Satarov, and
H.~St\"ocker for numerous discussions, and O. Lysak for the help
in the preparation of the manuscript. The work was supported in
part by US Civilian Research and Development Foundation (CRDF)
Cooperative Grants Program, Project Agreement UKP1-2613-KV-04, and
Virtual Institute on Strongly Interacting Matter (VI-146) of
Helmholtz Association, Germany.

\end{acknowledgments}

\appendix

\section{}\label{appendix-A}

The n-dimensional Gauss integral equals  the following
\cite{korn}:
 \eq{
 \int_{-\infty}^{+\infty}\cdots\int_{-\infty}^{+\infty}
 \exp\left[\,-\sum_{i,\,k=1}^n A_{i,k}\,x_ix_k
 \,\right]dx_1\ldots dx_n \;=\; \frac{\pi^{n/2}}{|A|^{1/2}}\;,
 }
where
%
 \eq{
 |A|
 \;\equiv\; \det A
 \;=\; \sum_{i=1}^n (-1)^{i+k} A_{i,k}\,M_{i,k}~,
 }
and
%
$ M_{i,k} \;=\; (-1)^{i+k} \partial |A|/\partial A_{i,k}$
%
is a complementary minor of the element $A_{i,k}$.
One also finds:
 \eq{
 \int_{-\infty}^{+\infty}\cdots\int_{-\infty}^{+\infty}
 x_ix_k\; \exp\left[\,-\sum_{i,\,k=1}^n A_{i,k}\,x_ix_k \,\right]dx_1\ldots dx_n
 \;=\; \frac{\pi^{n/2}}{|A|^{3/2}}\; (-1)^{i+k} M_{i,k} \;.
 }
%
%

%

%

\end{document}